\documentclass[a4paper,12pt]{article}
\usepackage[utf8x]{inputenc}
\usepackage{graphicx}
\usepackage{color}
\usepackage{authblk}
\usepackage{amsmath}
\usepackage{amssymb}
\usepackage{color}
\usepackage{textcomp}
\usepackage{caption}
\usepackage{hyperref}

\begin{document}
\title{\textbf{Non-monotonic Size Dependence of Diffusion and Levitation Effect: A Mode-Coupling Theory Analysis}}
\author{Manoj Kumar Nandi and Atreyee Banerjee and Sarika Maitra Bhattacharyya}
\affil{\textit{Department of Polymer Science and Engineering, National Chemical Laboratory, Pune-411008,India}}
\date{  }
\maketitle

\begin{center}
\textbf{Abstract}
\end{center}

We present a study of diffusion of small tagged particles in a solvent, using mode coupling theory (MCT) analysis and computer simulations. 
The study is carried out for various interaction potentials. For the first time, using MCT, it is shown that only for strongly attractive interaction 
potential with allowing interpenetration between the solute-solvent pair the diffusion exhibits a non-monotonic solute size dependence which has earlier 
been reported in simulation studies [\textcolor{black}{J. Phys. Chem. B \textbf{109}, 5824-5835 (2005)}].  For weak attractive and repulsive potential the solute size dependence of diffusion shows monotonic behaviour. It is 
also found that for systems where the interaction potential does not allow solute-solvent interpenetration, the solute cannot explore the neck of the solvent cage. 
Thus these systems  even with strong atrractive interaction will never show any non monotonic size dependence of diffusion. This non monotonic size dependence of 
diffusion has earlier been connected to  {\it levitation effect} [\textcolor{black}{J. Phys. Chem. \textbf{98},6368-6376 (1994)}].  We also show that although levitation 
is a dynamic phenomena, the effect of levitation can be obtained in the static radial distribution function. \\

\newpage

\section{Introduction}

 Diffusion phenomena in fluid has immense importance in various fields such as industrial chemistry, soft matter physics, and biology.
The motion of tagged particles in various mediums have been extensively studied using, microscopic theories, simulations and experiments [1-12].  These studies have predicted many interesting results. The theoretical studies of diffusion using mode coupling theory have predicted that the microscopic origin of many phenomena are quite different from what it had been envisaged earlier.
For example, the diffusion in a neat liquid (where the  solute particle is one of the solvent molecules) is usually found to follow the Stoke-Einstein relation (SER)  (\textbf{D}$\propto$1/\textbf{R} where \textbf{R} is the radius of the particle).  However, the SER should ideally be valid in the hydrodynamic regime where the solute particle is so big that it perceives the solvent as a continuum without any microscopic details. It was found that SER is valid because the viscosity
and the friction are determined by the same dynamical variable, namely the density auto correlation function \cite{review}\cite{sarika&biman2}.  Thus the study revealed that the validity of SER has a microscopic rather than a macroscopic hydrodynamic origin. 

The diffusion of small solute particles in a sea of larger solvents are known to violate the SER. Experiments have repeatedly shown that for small solute particles the SER significantly underestimates the diffusion coefficient\cite{pollack,evans} and this has been attributed to the micro viscosity effect or fractional
viscosity effect. Analytical MCT studies have shown that the decoupling of the solute motion from the solvent dynamics is the origin of the enhanced diffusion \cite{sarika&biman}.  The diffusion of small solutes not only show the breakdown of the SER, but also other anomalies. There has been a series of studies of diffusion of small solutes in different mediums by Yashonath and co-workers \cite{yash_mol}\cite{santi_jpc98}\cite{yashonath}\cite{yash144}\cite{yash145}. The medium in which diffusion 
was studied varied over a wide range from zeolites to other porous solids and also included Lennard-Jones (LJ) solvent. The simulation studies revealed that diffusion of small solutes show a diffusivity maximum as a function of the diffusant diameter, independent of the nature of the medium. This counter intuitive behaviour of solute diffusion through different types of porous structures was first reported by Yashonath and Santikary \cite{yash_mol}\cite{santi_jpc98} and later by Henson {\it et al} \cite{yash135}. Yashonath {\it et al} have systematically 
varied the size of the solute by keeping the porous zeolite structure the same \cite{santi_jpc98}\cite{yashonath}. 
In their study of diffusion through zeolite Y they have found that for a spherical sorbet of diameter 6.0$\AA$ which has dimensions  comparable to the bottleneck of the zeolite Y, the force exerted by the zeolite on the diffusing particle is minimum which 
leads to the maximum diffusion. In general they noticed that in zeolites  when the solute size is about 80$\%$ of the size of the neck of the zeolite the diffusion is maximum.
In their study of solute diffusion in LJ solvent they found that when the solute size is about 0.24 times that of the solvent the diffusion shows a maximum \cite{yashonath}. From Voronoi Polyhedra analysis they have obtained a void and neck size distribution of the solvent transient cages and have shown that, when the solute-solvent diameter is smaller but similar to the size of the neck, the solute can diffuse faster. They have attributed this diffusivity maximum to levitation effect (LE), stating that when the solute size and the size of the neck are similar then the attractive force felt by the solute while passing through the pore is uniform from all directions and this helps the particle levitate through 
the pore. Whereas, in case of smaller solutes, proximity to one part of the neck leads to a large asymmetry in the force felt by the solute. The solute tends to spend longer time bound to one side of the neck or one solvent particle (in case of LJ solvent), which leads to lower diffusion constant. In experiments, simulations and also in theoretical studies such maximum in diffusivity for ion diffusion in polar solvent has been found \cite{yash135}[13-16]\cite{yashonath2010}.  All these studies show that the non 
monotonic size dependence of diffusion is a generic phenomena independent of the nature of the solvent.

For understanding this behaviour observed in simulation we require a detailed knowledge of the diffusion mechanism. There has been earlier theoretical studies to understand the non monotonic solute size dependence of diffusion. 
Biswas {\it et.al} have used a self consistent microscopic theory to study ion diffusion in polar medium \cite{biswas}. They have shown that for certain solute sizes due to structure breaking of the solvent there is an enhanced diffusion which leads to the anomalous increase in diffusion of large ion in water. Sarkar {\it et.al} have written down a master equation to represent a sorbate diffusion in a zeolite \cite{sarkar}. They could connect the diffusion coefficient to the interaction potential felt by the sorbate. The theory could explain the levitation effect. Earlier studies have shown that mode coupling theory (MCT) can successfully explain diffusive dynamics and its anomalies over a wide range of solute-solvent systems\cite{review}. 
Hence it is imperative to study this non monotonic size dependence of diffusion in a solute-solvent system and its connection to levitation, using the MCT. 

In this article we use the fully self-consistent microscopic mode-coupling theory and molecular dynamic simulations to understand the origin of non-monotonic size dependence of diffusion and its relationship with levitation. 
Although the present study is confined only to solute diffusion through pure solvents, we expect that the understanding of the diffusion mechanism in this system will help us in future to study the solute diffusion in porous medium.
The present analysis involves four different systems by systematically varying the solute diameter for each system. The interactions between the solute
-solvent and  solvent-solvent pairs are modeled in terms of  LJ potential, where the interaction parameters are varied from large attractive to repulsive values. The
 solute-solvent radius $\sigma_{12}$ is varied in such a way that in some cases inter-penetration between the solute-solvent pair is allowed.  We show that for repulsive and weak attractive interaction the diffusion shows a monotonic size dependence whereas for strong attractive interaction the diffusion shows a non monotonic size dependence. For large attraction only certain solutes of intermediate sizes can levitate through the transient solvent cage leading to higher diffusion coefficient. The smaller solute particles cannot levitate and have larger probability of primarily sticking to one solvent molecule. 
The study also reveals that although levitation is a dynamic phenomena, its signature is present in the static radial distribution function (RDF).

The work is arranged as follows.
The simulation details are given in Sec.2. Sec.3 contains theoretical analysis. In Sec.4 the results are presented and finally, Sec.5 concludes with short discussion about the result.

\section{Simulation Details}

The atomistic models which are simulated are two component mixtures where particles of type i interact with those of type j with pair 
potentials, ($U_{ij}(r)$)  where r is the distance between the pairs. The potentials are shifted and truncated Lennard-Jones potentials,
\begin{equation}
 U_{ij}(r)=
\begin{cases}
 U_{ij}^{(LJ)}(r;\sigma_{ij},\epsilon_{ij})- U_{ij}^{(LJ)}(r^{(c)}_{ij};\sigma_{ij},\epsilon_{ij}),    & r\leq r^{(c)}_{ij}\\
   0,                                                                                       & r> r^{(c)}_{ij}
\end{cases}
\end{equation}
where $U_{ij}^{(LJ)}(r;\sigma_{ij},\epsilon_{ij})=4\epsilon_{ij}[({\sigma_{ij}}/{r})^{12}-({\sigma_{ij}}/{r})^{6}]$
 ,$r^{(c)}_{ij}=2^{\frac{1}{6}} \sigma_{ij}$ for Week Chandler Anderson (WCA) system \cite{weeks}  and
$r^{(c)}_{ij}=3.5\sigma_{ij}$ for LJ system. Where i,j=1,2. 1 refers to solvent and 2 refers to solute.

Four systems, consisting 456 solvent or host atoms ($N_{h}$) and 44 solute 
or guest atoms ($N_{g}$)  are studied.
$N_{h}+N_{g}=500$.

The different models are distinguished by different choices of lengths and energy parameters. Lengths, temperature and
time are given in units of $\sigma_{11}$, ${k_{B}T}/{\epsilon_{11}}$ and $\surd({m\sigma_{11}^2}/{\epsilon_{11}})$ 
respectively.
The details of the models are given in the Table (I) and $m_{1}=m_{2}=m=1$ for all the cases. Note that for system 1, 2 and 3, the interaction between the solute and the solvent has a soft core which allows inter-penetration between solute solvent pair. However in system 4 the solute-solvent diameter is given by Lorentz-Berthelot (LB) rule \cite{LB-rule}
\begin{table}
\caption{Parameters for model systems used in this study, where $\sigma_{11}$, $\sigma_{22}$ and $\sigma_{12}$ are the solvent, solute and solute-solvent diameter respectively and 
$\epsilon_{11}$, $\epsilon_{22}$ and $\epsilon_{12}$ are the interaction parameters for the solvent-solvent, solute-solute and solute-solvent pairs, respectively. For system 1, 2 and 3 inter-penetration between the solute-solvent pair is allowed
and system 4 follows LB rule for mixing and the $\sigma_{22}$ values are 0.073, 0.098, 0.122, 0.171, 0.200, 0.220, 0.244, 0.293, 0.317, 0.366.}
\begin{tabular}{||l|r|r|r|r|r|r|r|}
 \hline
$System-Potential$   & $\sigma_{11}$&$\sigma_{12}$ &$\sigma_{22}$ & $\epsilon_{11}$ &$\epsilon_{22}$&$\epsilon_{12}$\\
 \hline
  
1-LJ & 1&${\sigma_{22}+0.171}$ & from & 1.0 & 3.96 & 6.0 \\
2-LJ & 1&${\sigma_{22}+0.171}$ & 0.073 & 1.0 & 1.0 & 1.0 \\
3-WCA & 1&${\sigma_{22}+0.171}$ & to & 1.0 & 3.96 & 6.0 \\
4-LJ & 1&$({\sigma_{11}+\sigma_{22}})/{2}$ & 0.366 & 1.0 & 3.96&6.0 \\
\hline
\end{tabular}

\end{table}

If $\epsilon_{11}$=0.25 KJ/mole,and $\sigma_{11}$=4.1 $\AA$, $\sigma_{22}$ is varied between 0.3 $\AA$ to 1.5 $\AA$ ,$m_{1}=40 amu$, then
$\tau$=5.186 ps and 
the system is identical with set-4 studied by Yashonath et.al \cite{yashonath}.

The sample is cubic with side length 7.86789$\sigma_{11}$.
We have carried out MD simulations at reduced temperature, T*=1.663,
reduced density for the solvent, $\rho*$=0.933,in micro-canonical ensemble(NVE) using velocity integration with time 
step of 0.001$\tau$.

For all the simulations mentioned above,the system has been equilibrated for 1 ns followed
by production run of 2 ns. The molecular dynamics simulations carried out using LAMMPS 
package \cite{lammps}.

The simulation results have also been checked for $N_{g}$=1 and the results (like RDF and D) remain identical with that for $N_g$=44.

\section{Theoretical Analysis}
 
For the mode-coupling calculation we consider a single solute (guest) particle in a sea of solvent (host) particles. The diffusion coefficient(D) of the 
tagged solute is given by the well known Einstein relation \\
\begin{equation}
\label{eins}
D=k_{B}T/m\Gamma(z=0) 
\end{equation}
 Where $\Gamma(z)$ is the frequency dependent friction on the particle having mass $\textquoteleft{m}$'. The frequency dependent friction $\Gamma(z)$ can be calculated using mode coupling theory. The
separation of time scale between the binary and repeated recollisions are invoked to decompose the friction into short and long time part. The short time part arises due to direct
 binary collisions,  and a long time part which arises due to correlated recollisions.  
The calculation of the recollision term is non trivial and includes information of the five hydrodynamic modes.  It has been reported earlier that for small solute particles, the 
density mode of the solvent primarily contributes to the recollisional friction. If we neglect the contribution from other hydrodynamic modes then the friction can be written as, 
\cite{sarika&biman}
\begin{eqnarray}
\label{fricforcalculation}
\Gamma^{ij}(z)\simeq\Gamma^{ij}_{B}(z)+R^{ij}_{\rho\rho}(z)
\end{eqnarray}
where $\Gamma^{ij}_{B}(z)$ is binary collisional friction. $R^{ij}_{\rho\rho}(z)$  arises from the coupling
of the tagged particle motion with the density fluctuation of the medium.

A single binary collision between the tagged particle and a solvent particle , in presence of the other solvent particles is described by $\Gamma_B$. The calculation of the full time 
dependence of $\Gamma_B$ is non trivial. However it has been shown that the re collision term begins as $t^{6}$ \cite{sjogren&sjolender}, thus the binary collision term contains all
 the contributions to the order $t^{2}$. As only even powers of $t$ appears thus the binary collision can be assumed to be a Gaussian,
\begin{equation}
\label{tau}
 \Gamma^{ij}_{B}(t)=\omega_{0ij}^2exp(-t^2/{\tau_{ij}}^2)
\end{equation}
here $\omega_{0ij}$ is the Einstein frequency :
\begin{equation}
\label{einfric}
 \Gamma^{ij}_{B}(t=0)=\omega_{0ij}^2=\frac{\rho}{3m}\int{d\textbf{r}g_{ij}(\textbf{r})\nabla^2v_{ij}(\textbf{r})}
\end{equation}
where $v_{ij}(r)$ is the inter-particle potential and $g_{ij}(r)$ is the static pair correlation 
function (RDF) between the $i$ and the $j$ type particles, $\rho$ is the particle number density. In Eq. \ref{tau} the 
relaxation time $\tau_{ij}$ is determined by taking the second derivative of a more general expression for binary friction and is given 
by \cite{sjogren&sjolender},
\begin{eqnarray}
\label{timecons}
 \omega_{0ij}^2/\tau_{ij}^2=(\rho/3m^2)\int{d\textbf{r}(\nabla^{\alpha}\nabla^{\beta}v_{ij}(\textbf{r}))g_{ij}(\textbf{r})(\nabla^{\alpha}\nabla^{\beta}v_{ij}(\textbf{r}))}\nonumber\\
  +(1/6\rho)\int{[d\textbf{q}/(2\pi)^3]\gamma_{dij}^{\alpha\beta}(\textbf{q})(S_{11}(\textbf{q})-1)\gamma_{dij}^{\alpha\beta}(\textbf{q})}
\end{eqnarray}
where summation over repeated indices is implied. $S_{11}(\textbf{q})$ is the static structure factor for the pure solvent and $\gamma_{dij}^{\alpha\beta}$ is given by

\begin{eqnarray}
 \gamma_{dij}^{\alpha\beta}(\textbf{q})=-(\rho/m)\int{d\textbf{r}exp(-i\textbf{q.r})g_{ij}(\textbf{r})\nabla^{\alpha}\nabla^{\beta}v_{ij}(\textbf{r})}\nonumber\\
=\hat{q}^{\alpha}\hat{q}^{\beta}\gamma_{dij}^l(\textbf{q})+(\delta_{\alpha\beta}-\hat{q}^{\alpha}\hat{q}^{\beta})\gamma_{dij}^t(\textbf{q})
\end{eqnarray}

where $\gamma_{dij}^l(\textbf{q})=\gamma_{dij}^{zz}(\textbf{q})$ and $\gamma_{dij}^t(\textbf{q})=\gamma_{dij}^{xx}(\textbf{q})$\\

The term arising in the friction due to the coupling to the density fluctuations in the surrounding medium is given by
\begin{eqnarray}
\label{rrhorho}
 R^{ij}_{\rho\rho}(t)&=&(\rho/m\beta)\int[{d\textbf{q}^{\prime}}/(2\pi)^3](\hat{\textbf{q}}.\hat{\textbf{q}}^{\prime})^2{q'}^{2}S_{11}(q)\nonumber\\
             & & {} \times  {[c_{ij}(q')]}^2[\phi^s(q',t)\phi(q',t)-\phi_0^{s}(q',t)\phi_0(q',t)]
\end{eqnarray}
Where $\beta$=${1}/{k_{B}T}$ and $\hat{\textbf{q}}^{\prime}$ is an unit vector along the arbitrary z direction and $c_{ij}(q)$
is the two particle direct correlation function between particles $i$ and $j$. 
$\phi(q,t)$ and $\phi^{s}(q,t)$ are the full and the self intermediate scattering functions respectively. $\phi_0(q,t)$=exp(-$({q^2t^2}/{2m{\beta}S_{11}(q)})$), 
and $\phi^{s}_{0}(q,t)=exp(-({q^2t^2}/{2m\beta}))$ are the inertial part of the intermediate scattering function and it's self part, respectively.

$\phi_{s}(q,t)$ and $\phi(q,t)$ are calculated self consistently using the well known MCT equations (\ref{mct1} ,\ref{mct2}). 
\begin{equation}
\label{mct1}
 \ddot{\phi}_{q}(t)+{\Gamma^B}_{11}(q,z=0)\dot{\phi}_{q}(t)+\Omega_{q}^2\phi_{q}(t)+\int_0^t\mathcal{M}_{qij}(t-\tau)\dot{\phi}_{q}(\tau)d\tau =0
\end{equation}
\begin{equation}
\label{mct2}
 \ddot{\phi^s}_{q}(t)+{\Gamma^B}_{ij}(q,z=0)\dot{\phi^s}_{q}(t)+\Omega_{0}^2{\phi^s}_{q}(t)+\int_0^t\mathcal{M}_{qij}^{s}(t-\tau)\dot{\phi^s}_{q}(\tau)d\tau =0
\end{equation}
where $\Omega_{q}^2=({q^2k_{B}T}/{mS_{11}(q)})$ and $\Omega_{0}^2=({q^2k_{B}T}/{m})$ and $\mathcal{M}_{qij}(t-\tau)$ and $\mathcal{M}_{qij}^{s}(t-\tau)$ are the memory kernels. They give rise to the long-time tails in the intermediate scattering function and its self part. The short time part is considered to be delta-correlated with the strength given by $\Gamma^B_{ij}(q,z=0) $. $\Gamma^B_{ij}$(q,z=0) is written as
\begin{equation}
 \Gamma^B_{ij}(q,z=0)=\int{dt\omega^2_{0ij}exp(-t^2/\tau^2_{ij}(q))}
\end{equation}
Where $\tau_{ij}(q)$ is calculated from  
\begin{eqnarray}
\omega^2_{0ij}/\tau^2_{ij}(q)=\frac{5q^2\omega^2_{0ij}}{2m\beta}+(\rho/3m^2)\int{d\textbf{r}(\nabla^{\alpha}\nabla^{\beta}v_{ij}(\textbf{r}))g_{ij}(\textbf{r})(\nabla^{\alpha}\nabla^{\beta}v_{ij}(\textbf{r}))}\nonumber\\
 +(1/6\rho)\int{[d\textbf{q}/(2\pi)^3]\gamma_{dij}^{\alpha\beta}(\textbf{q})(S_{11}(\textbf{q})-1)\gamma_{dij}^{\alpha\beta}(\textbf{q})}
 \end{eqnarray}
 
\begin{eqnarray}
\label{phimemory}
\mathcal{M}_{q11}(t)&=&\frac{k_BT\rho}{16\pi^3}\int d{\bf q}\left[\textbf{k}\cdot({\bf q}c_{11}(q)+({\textbf k}-{\bf q})c_{11}(k-q))\right]^2S_{11}(q)S_{11}(k-q)\nonumber\\
                            && {} [\phi(q,t)\phi(q-k,t)-\phi_{0}(q,t)\phi_0(q-k,t)]
\end{eqnarray}

\begin{equation}
\label{fsmemory}
 \mathcal{M}_{qij}^s(t)=\frac{k_BT\rho}{(2\pi)^{3}}\int{d\textbf{q}S_{11}(q)c_{ij}(q)^2(\frac{\textbf{q.k}}{q^2})^2[\phi_{s}(q-k,t)\phi(q,t)
-\phi_{0}^{s}(q-k,t)\phi_0(q,t)]}
\end{equation}

Eq(\ref{phimemory}) is the memory function for the $\phi(q,t)$ calculation and Eq(\ref{fsmemory}) is for $\phi_{s}(q,t)$

The integration in the memory-functions are converted to Riemann sums for the discrete momentum values \cite{fuch}. We take 500 grid points 
with q starting from 0.2 upto a cutoff of 199.8 with step size dq=0.4. Usually MCT calculations are done with q cutoff of 39.8 but for our
calculation since the solute sizes are small we need to go until $q=199.8$.

We use $g_{11}(r)$ and $g_{12}(r)$ from theory and simulation and $S_{11}(q)$ and $c_{11}(q)$ and $c_{12}(q)$ are calculated analytically using HMSA (Hypernetted Mean Spherical Approximation) closer \cite {hansen}.
We get $\phi(q,t)$ and $\phi_{s}(q,t)$ by solving  Eqs.(\ref{mct1} and \ref{mct2}) by self consistent method.

\section{Result and Discussion}
 The Mode coupling Theory (MCT) formalism discussed above has been employed to calculate the self diffusion constants, $D$. The diffusion value for a pure solvent using this fully self consistent MCT at $T^{*}=1.663$ and $\rho^{*}=0.933$, has been found to be $D_{MCT}=0.194\times10^{-4} cm^{2}/s$ which is close to the simulated value $D_{sim}=0.195\times10^{-4} cm^{2}/s$. 
 
\begin{figure}[h]
\centering
\includegraphics[width=8cm,height=6cm]{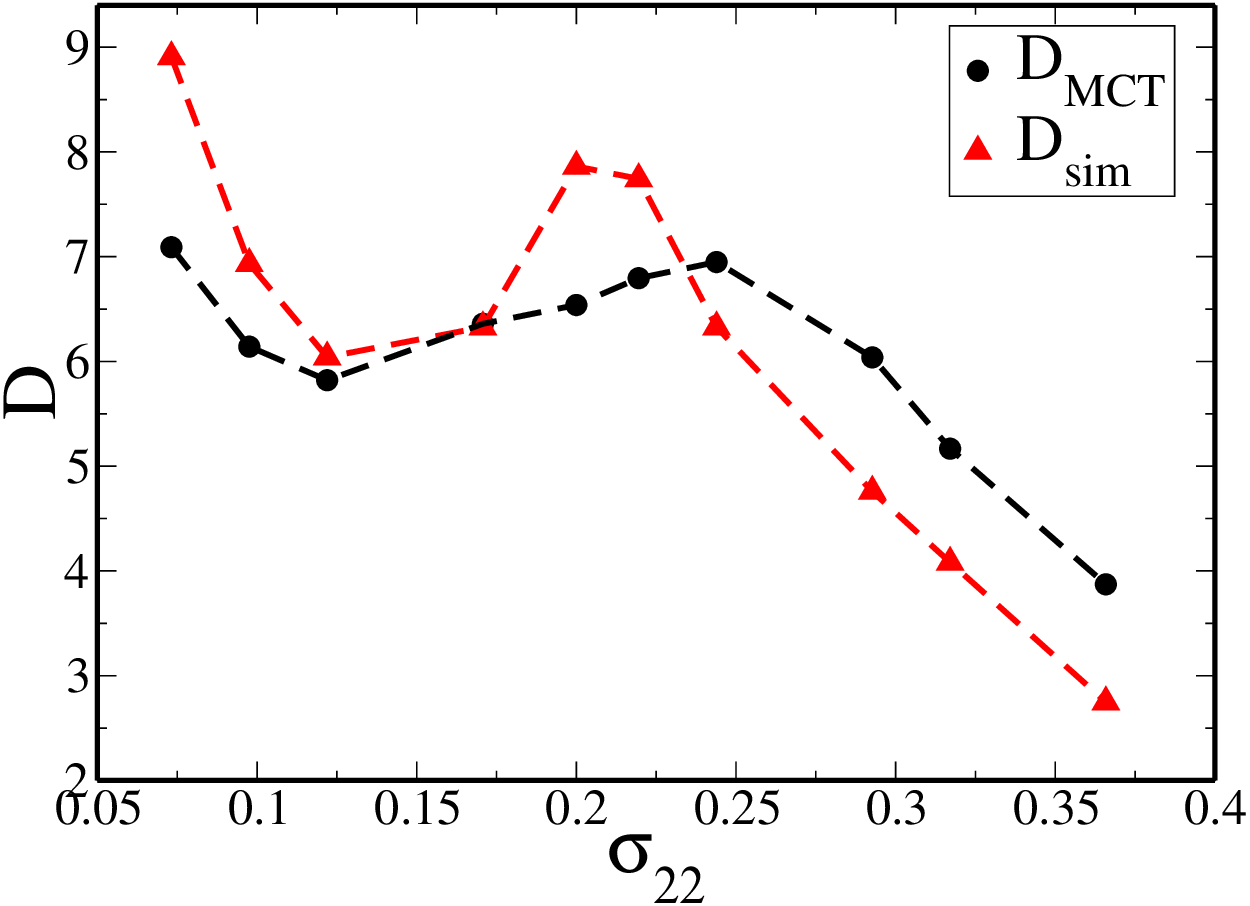}
\caption{Scaled diffusion coefficients obtained from fully self consistent MCT, $D_{MCT}$ (circles) and from simulations, $D_{sim}$ (triangles) are plotted against $\sigma_{22}$ for system 1.  Both show non-monotonic size dependence of diffusion with a maximum diffusion. To plot both in the same graph, the diffusion values are scaled by arbitrary constants. The $\sigma_{22}$ values used for the calculation are 0.073,0.098,0.122,0.171,0.200,0.220,0.244,
0.293,0.317 and 0.366}
\label{fig1}
\end{figure} 

 In Fig.\ref{fig1} we present both simulated and calculated (using full self-consistent MCT)  diffusion values against the solute diameter for system 1 (see Table I). The studies show a non monotonic size dependence of diffusion. The theory and simulation studies both predict that as the solute size is decreased the diffusion value first increases and then decreases. This diffusivity maximum as discussed earlier has been attributed to levitation effect \cite{yashonath}. Thus it is found that MCT can predict levitation effect and diffusivity maximum as found in simulations \cite{yashonath} and experimental \cite{yashonath2010} studies. 
 
 Although the nature of diffusion as a function of the solute diameter, predicted by the fully self consistent MCT, is similar to that found in simulation studies, the value of diffusion obtained from this method is found to be much smaller than that obtained from simulations. Note that in Fig.\ref{fig1} the $D_{sim}$ and $D_{MCT}$ are scaled by arbitrary values. Thus, although the diffusion value for a pure solvent obtained using fully self consistent MCT, matches with that of the simulated results, for smaller solutes the fully self-consistent MCT seems to overestimate the friction.  Compared to the simulation values, the self intermediate scattering function, $\phi_{s}(q,t)$  obtained using this fully self-consistent 
method (from Eqs.\ref{mct1} and \ref{mct2}) is found to have a longtime tail. For smaller solutes this discrepancy between simulated and analytical $\phi^{s}$ increase in the long time. This implies that the wavenumber dependent friction, 
$\mathcal{M}^{s}_{qij}$, which is the memory function for $\phi^{s}$ and determines its longtime tail is not the correct approximation in the limit of small solute particles. There should be other terms in the memory function which should lower the value of the wavenumber dependent friction, $\mathcal{M}^{s}_{qij}$, and thus the long time tail. 

 To avoid this problem we have calculated  approximate $\phi^{s}$ using simulated mean square displacement $<\Delta r^{2}(t)>$ and non Gaussian parameter $\alpha_{2}=({3<\Delta r^{4}(t)>}/{5<\Delta r^{2}(t)>^{2}})-1$.  The  $\phi^{s}_{q}$  expression is given by \cite {hansen-and-mcdonal}, 

\begin{equation}
\label{non-gaussian}
\phi^{s}_{q}(t)=exp\left(-\frac{q^{2}<\Delta r^{2}(t)>}{6}\right)
\left[1+\frac{1}{2} \alpha_{2} \left(\frac{q^{2}<\Delta r^{2}(t)>}{6}\right)^{2}\right]
\end{equation}
\noindent

 \begin{figure}[h]
\centering
\includegraphics[width=8cm,height=6cm]{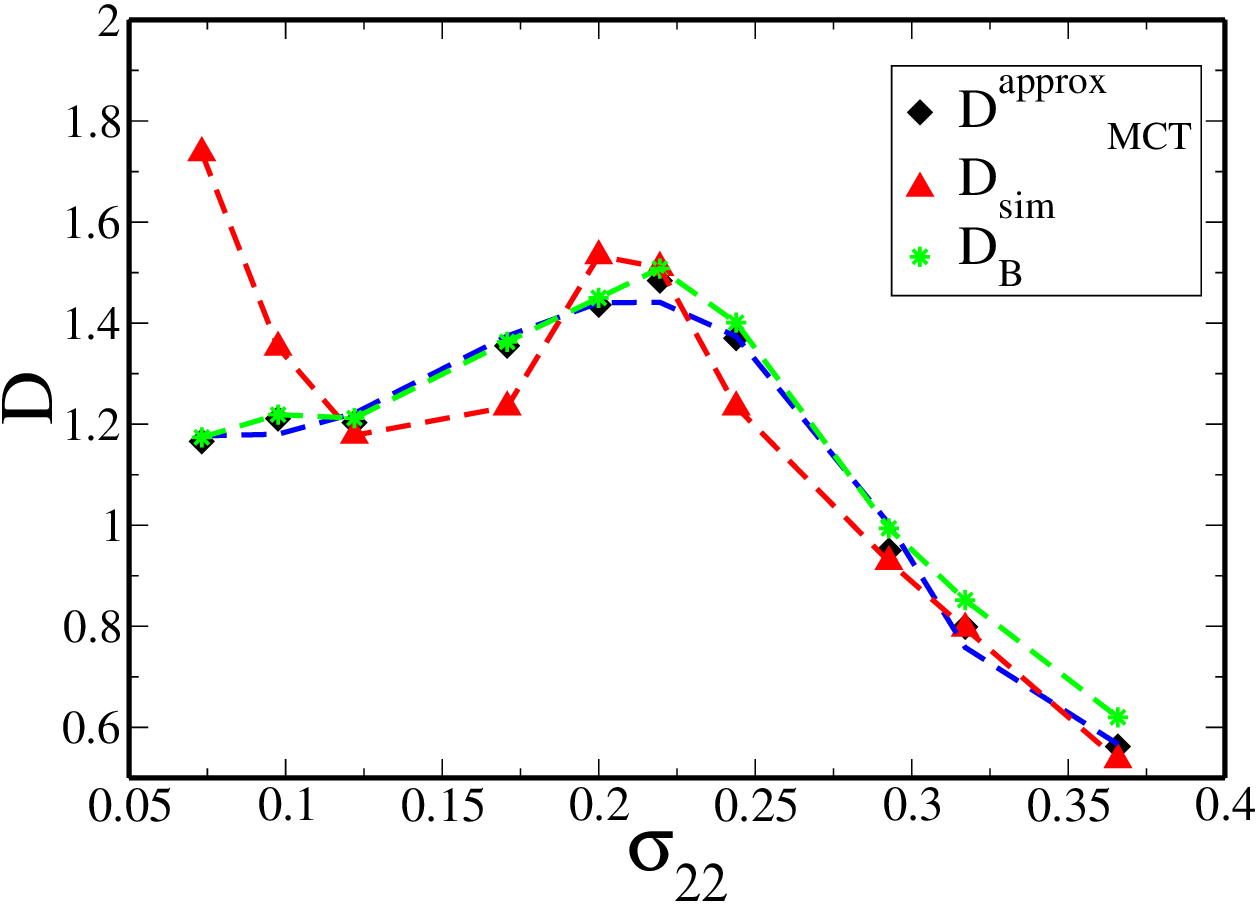}
\caption{Diffusion coefficients obtained from MCT where $\phi_{s}(q,t)$ is calculated using Eq.\ref{non-gaussian}, $D^{approx}_{MCT}$ (diamonds), and from simulations, $D_{sim}$ (triangles) are plotted against  $\sigma_{22}$ for system 1. 
We also plot the bare diffusion coefficient $D_{B}$ (stars). Diffusion values are scaled by $10^{-4}cm^2/s$.  
The solute diameters used for this calculation are same as in Fig.\ref{fig1}. The diffusion values show non monotonic size dependence.}
\label{fig2}
\end{figure}

With this approximation for the self intermediate scattering function the diffusion/friction has been calculated using Eqs. 2-8. Note that  the dynamic structure factor is always calculated self-consistently. In the calculation the simulated radial distribution function has been used 
as the effect of levitation is stronger in the simulated RDF. This point will be discussed later.  In Fig.\ref{fig2} the diffusion values calculated using approximate $\phi^{s}_{q}(t)$(from Eq.\ref{non-gaussian}), $D^{approx}_{MCT}$, are plotted against the solute diameter for system 1. For comparison the simulated values, $D_{sim}$ are also plotted.  $D_{approx}$ values are similar to $D_{sim}$ values and does show a non monotonic size dependence.  However, the peak value is little shifted and for the two smallest sizes (0.073 and 0.098) the diffusion values are smaller than that predicted in simulation studies. In the same figure the bare diffusion coefficients obtained from inverting the binary frictions, $(D_{B}=({k_{B}T}/{m\Gamma_{B}}))$ are also plotted. It is found that for small solute sizes the binary contribution to friction is dominant and the contribution from the density term $R^{ij}_{\rho\rho}$ is negligible. The frictional contributions from the other hydrodynamic modes like the longitudinal and transverse current \cite{sjogren}  have also been calculated and it was found that none of the hydrodynamic modes actually contribute to the friction/ diffusion. For small solutes as found earlier\cite{sarika&biman},  the solute motion is decoupled from the solvent dynamics and  the hydrodynamic modes of the solvent do not play any role in the solute dynamics. {\it For these small sizes there may be some contribution from non-hydrodynamic modes which assist in the particle diffusion.}

Except for the smallest two sizes the behaviour of diffusion constant as predicted by the analytical MCT calculation is similar to that predicted by  the simulation studies.  As mentioned in the Introduction, this anomalous size dependence of diffusion has been attributed to levitation effect \cite{yashonath} where the solute particles with intermediate sizes are expected to levitate through the neck of the solvent cage whereas the smaller particles are expected to be strongly attracted to one solvent particle.

\begin{figure}[h]
\centering
\includegraphics[width=8cm,height=6cm]{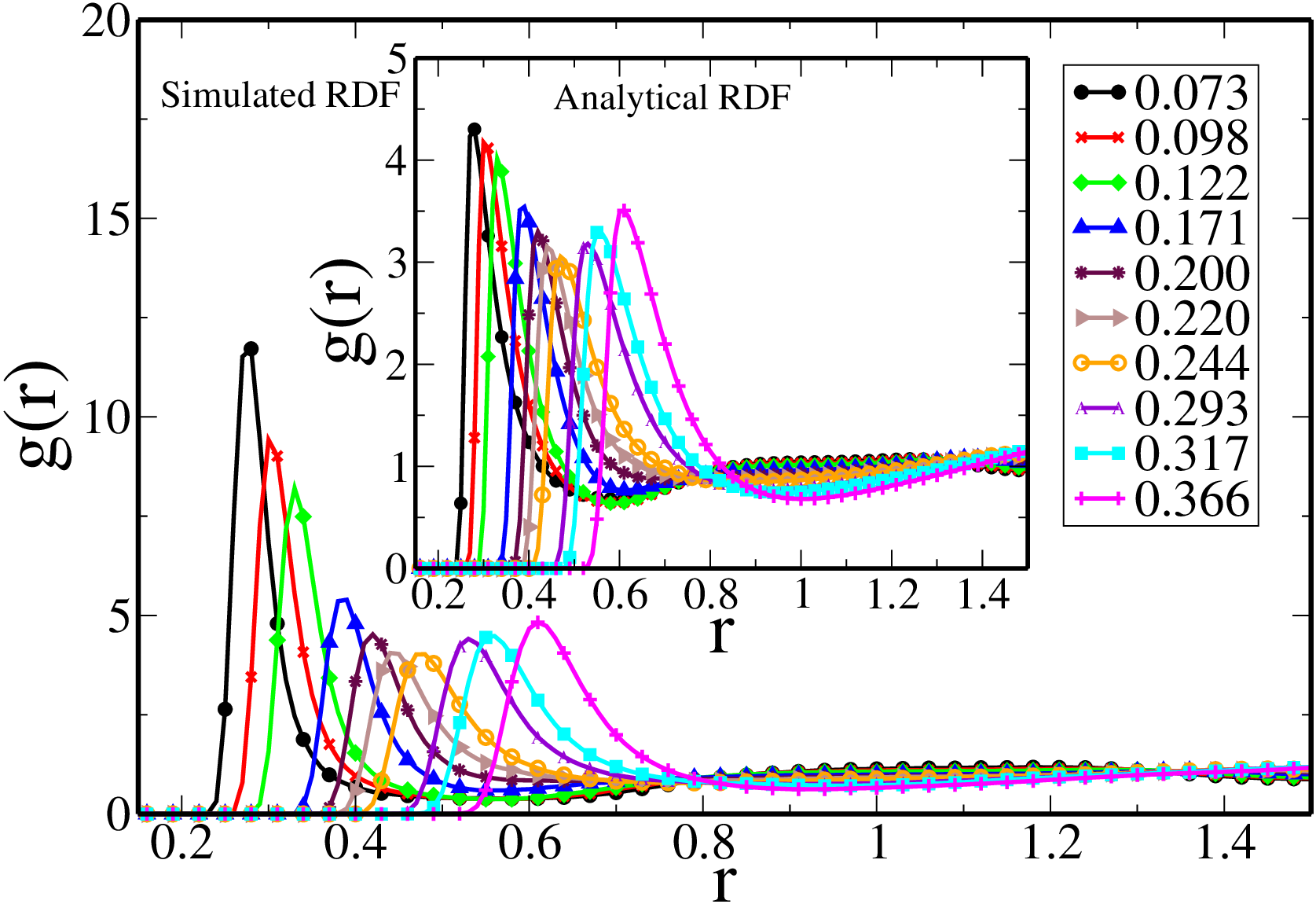}
\caption{Radial distribution functions from simulation are plotted for different solute sizes where the solute sizes are varied as given in Fig.\ref{fig1}. In the inset the same is plotted as obtained using HMSA closer \cite{hansen}. For both the sets of RDFs the peak value of the maxima show a non-monotonic size dependence. The effect is stronger in the simulated RDF. }
\label{fig3}
\end{figure}

The effect of the above mentioned phenomena is reflected in the solute-solvent radial distribution function. In Fig.\ref{fig3} it is shown that the peak value of the RDF does show a non monotonic size dependence of diffusion as a function of solute size. The RDFs for the intermediate solute sizes have a lower and broader diffused 1st peak, whereas the value of the 1st peak of the RDFs for smaller solutes are higher and the peaks are also much sharper. We also note the non monotonicity is more prominent in the simulated RDF. However, the analytical RDF also shows a weak non monotonic behaviour thus predicting that the HMSA closer can show a weak effect of the levitation.  
The effect of levitation on RDF will be discussed in details later in this section.    
As discussed earlier, for this range of solute sizes, the binary contribution to the friction is dominant.  The binary friction (Eq. 4-7) is primarily determined by the RDF. Thus the effect of levitation which is manifested in the RDF does finally give rise to the diffusion anomaly. 

For system 1 where  the anomalous non-monotonic size dependence of diffusion is observed, the particles interact via LJ potential and there is a strong attraction between the solute-solvent pair ($\epsilon_{12}=6.0$). Note that the solute-solvent diameters in this system are smaller than that predicted by Lorentz-Berthelot rule \cite{LB-rule}. Thus for this system the interaction potentials have soft core and interpenetration between the solute and the solvent pairs are allowed.

To find out the individual roles of the solute-solvent interpenetration and strong attraction play in the anomalous behaviour of the diffusion, we further study three more systems(see Table I).  For system 1, 2 and 3 that interpenetration is allowed, however the attractive part of the interaction for system 2 is weaker than that for system 1 and  for system 3 the interaction is kept only repulsive. Between system 1 and 4 , both have strong attractive interaction but system 4 does not allow any interpenetration between solute-solvent. 
\begin{figure}[h]
\centering
\includegraphics[width=12cm,height=6cm]{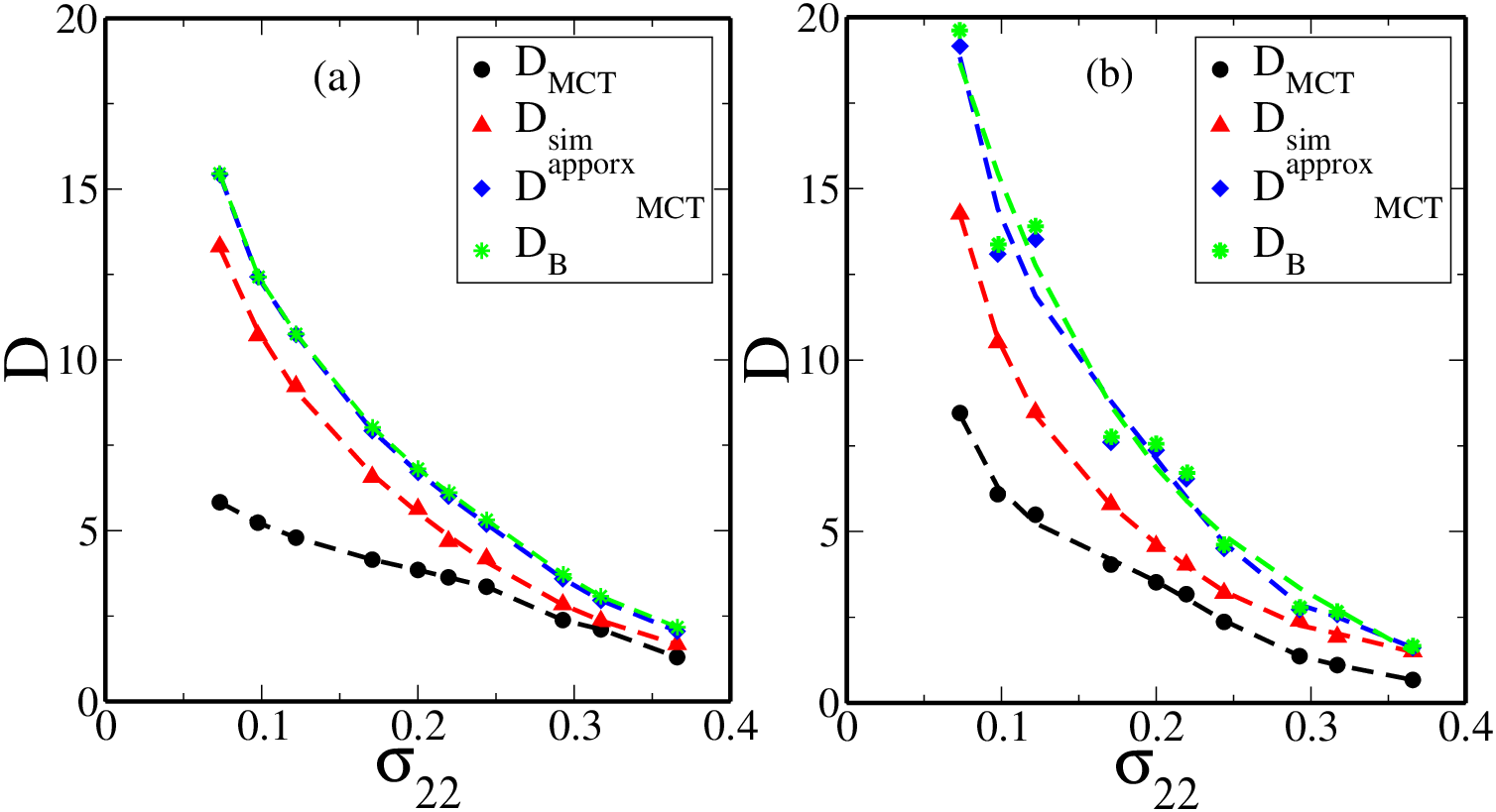}
\caption{(a) Diffusion coefficients obtained from fully self consistent MCT, $D_{MCT}$ (circles), from MCT where $\phi_{s}(q,t)$ is calculated using Eq. \ref{non-gaussian}, $D^{approx}_{MCT}$ (diamonds), and from simulations, $D_{sim}$ (triangles) are plotted against  $\sigma_{22}$ for system 2. We also plot the bare diffusion coefficient $D_{B}$ (stars). (b) Same as in (a) but for system 3.
Diffusion values are scaled by $10^{-4}cm^2/s$.
Non of the diffusion values show any non monotonic solute size dependence. The solute values are same as in Fig.\ref{fig1}.}
\label{fig4}
\end{figure}

\begin{figure}[h]
\centering
\includegraphics[width=12cm,height=6cm]{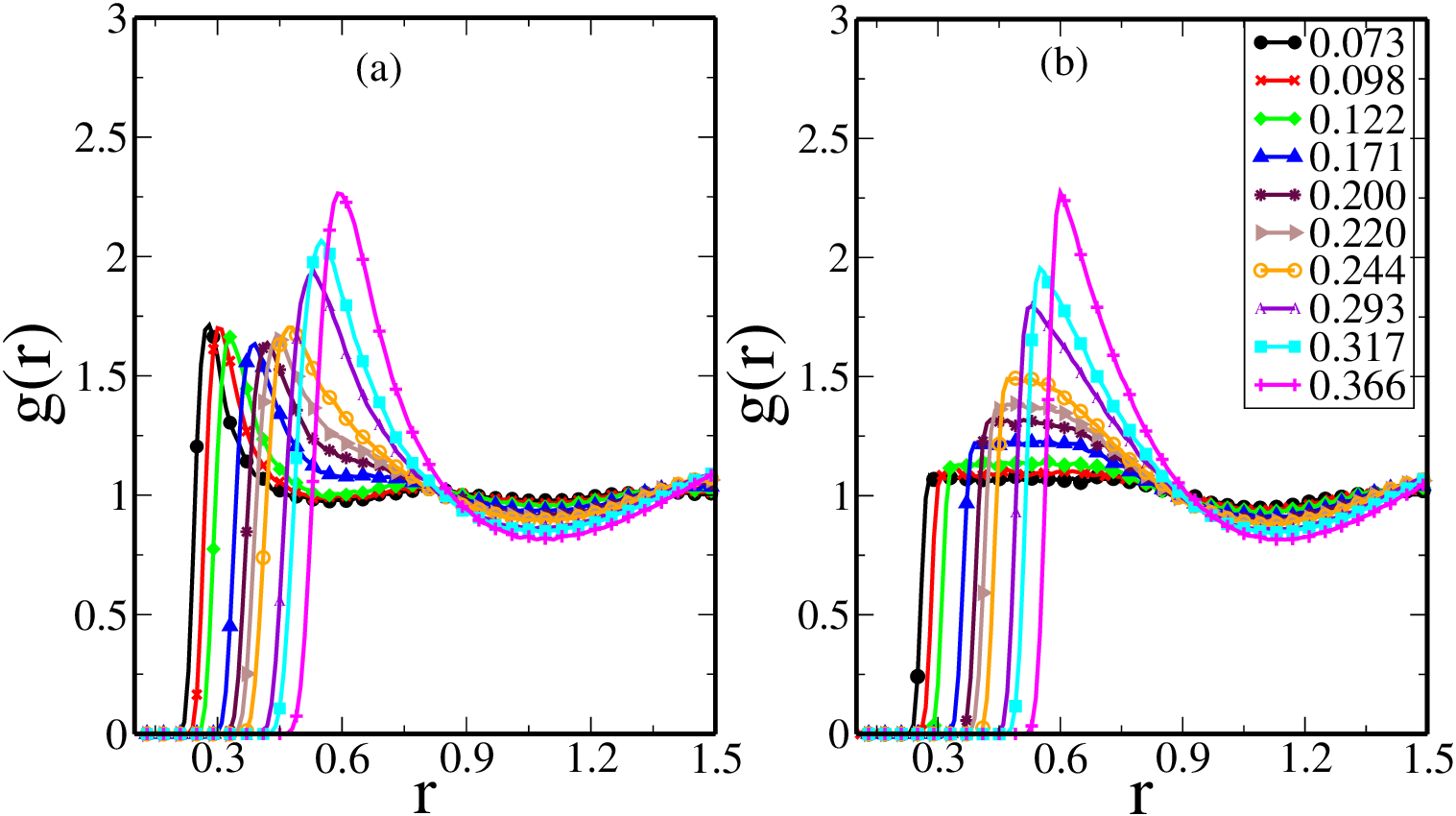}
\caption{Radial distribution function (RDF) for system 2 (a)  and system 3 (b)  for all solute particles. The peak values of the RDFs monotonically decrease with the solute size. For system 3 which is purely repulsive, the RDF is almost structure less for small sizes.}
\label{fig5}
\end{figure}
The diffusion values, $D_{MCT}$, $D^{approx}_{MCT}$, $D_{sim}$ and $D_{B}$ for systems 2 and 3 are shown in Fig.\ref{fig4}.  Neither simulation nor theory show any non monotonic size dependence of diffusion. This absence of diffusivity maximum in absence of attractive interaction was earlier found in simulation studies of sorbate diffusion in zeolites \cite{santi_jpc98}. 
The complete self consistent theory does show a small maximum which comes from the density contribution (Eq.\ref{rrhorho}). However, this kink can be an artifact of the full self consistency which disappears when the approximate self dynamic structure factor obtained using inputs from simulations (Eq. \ref{non-gaussian}) is used to calculate the density contribution. 
Just like the diffusion coefficients, the 1st peak values of the radial distribution function in Fig.\ref{fig5} also does not show any strong non monotonic behaviour. For system 2 a slight non-monotonicity is observed but for system 3 it is completely monotonic. Similar to system 1, in system 2 and 3, the friction is primarily determined by the binary collisions, which in turn depends on the RDF. Thus non-monotonicity in diffusion coefficient implies that the RDF will be non-monotonic, however, the reverse might not be always true.
From this analysis it can be concluded that only in presence of {\it strong } attractive interaction between the solute-solvent pair, the diffusion shows a non monotonic solute size dependence. 

\begin{figure}[h]
\centering
\includegraphics[width=12cm,height=6cm]{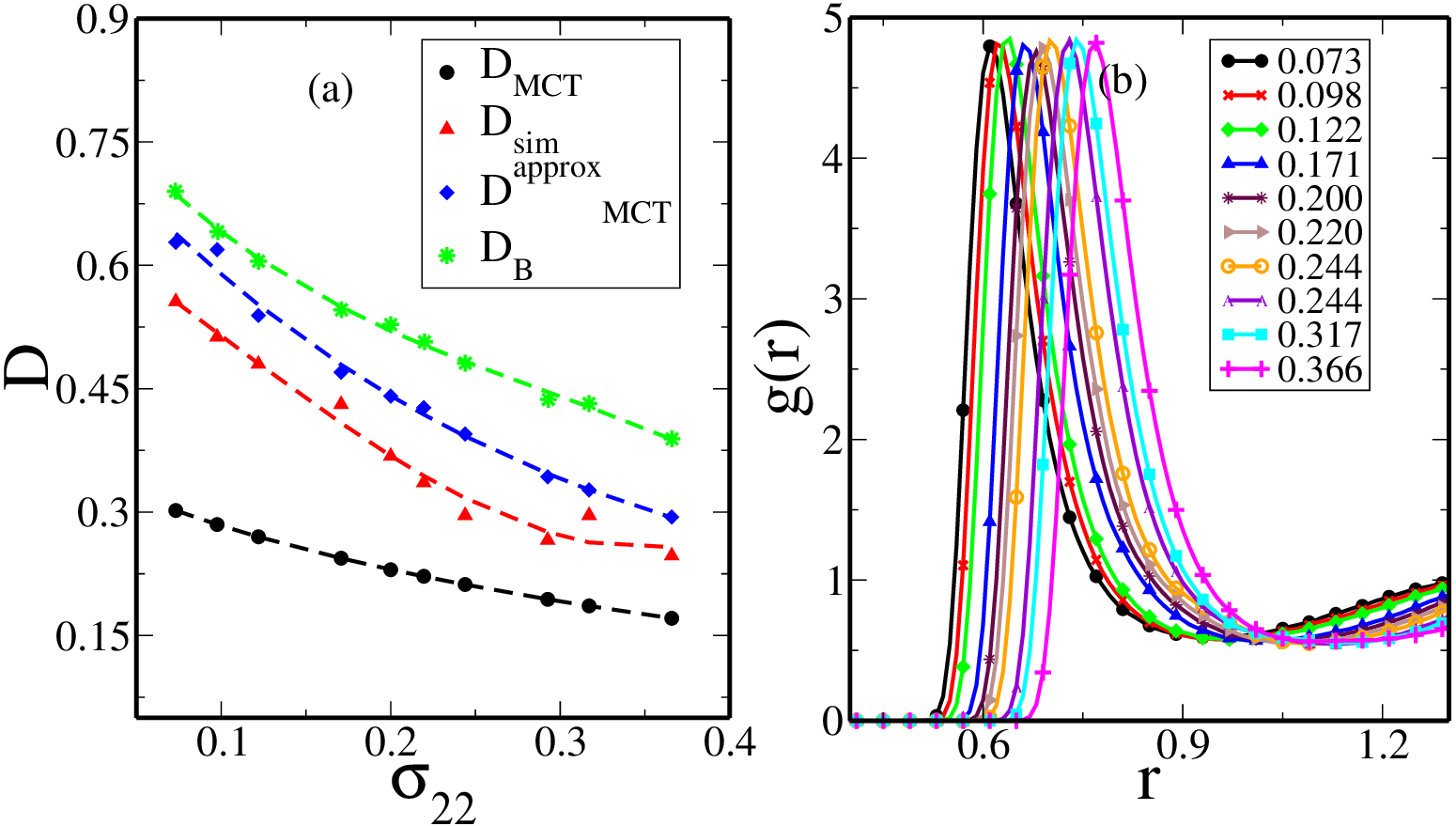}
\caption{(a) Diffusion coefficients obtained from fully self consistent MCT, $D_{MCT}$ (circles), from MCT where $\phi_{s}(q,t)$ is calculated using Eq.\ref{non-gaussian}, $D^{approx}_{MCT}$ (diamonds), and from simulations, $D_{sim}$(triangles) are plotted against $\sigma_{22}$ for system 4. We also plot the bare diffusion coefficient $D_{B}$ (stars). Diffusion values are scaled by $10^{-4}cm^2/s$. The diffusion values does not show any non monotonic behaviour as a function of solute size. The solute diameters are same as in Fig.\ref{fig1}. (b)Radial distribution function (RDF) for system 4 for all solute particles.  This plot shows well defined RDF with no non monotonic solute size dependence of the peak value.}
\label{fig6}
\end{figure}

 In order to understand the role of solute-solvent interpenetration in this anomalous behaviour of the diffusion we simulate a system where the diameter for the solute-solvent pair is given by,  $\sigma_{12}=({\sigma_{11}+\sigma_{22}})/{2}$, following Lorentz-Berthelot rule. The interaction parameter for this system has been kept same as that for system 1, i.e the solute-solvent interaction has been kept strongly attractive ($\epsilon_{12}=6.0$). As shown in Fig.\ref{fig6}a, even though we have strong attraction between the solute and the solvent, without the interpenetration the diffusivity maximum is not present. The RDF for this system as shown in Fig.\ref{fig6}b does not show any non monotonic behaviour. 
 Unlike in systems 1,2 and 3, in system 4 the friction is determined by both the binary and the density contribution terms.  

For the solvent studied here we know that the neck size distribution of the transient solvent cage as obtained from the Voronoi Polyhedra analysis has a peak around $\sigma_{neck}=0.425\sigma_{11}$  (see figure 5(b) of Ref. \cite{yashonath}).  Yashonath {\it et al.}  in all their studies have defined a levitation parameter $\gamma=({2^{1/6} \sigma_{12}}/{\sigma_{neck}})$ \cite{yashonath} \cite{yasho11}. Usually for zeolitic system they have found that when the levitation parameter, $\gamma$ is close to unity the diffusion reaches a maximum value. For solute-solvent system they have found that the diffusion value is maximum when $\gamma \approx 0.62$. We find that for systems which follow LB rule even for the smallest solute sizes,  the levitation parameter is always greater than unity. Thus these systems will never predict any anomalous non monotonic size dependence of diffusion.

 In all the systems it is found that $D_{MCT}$ underestimates the diffusion value, however, $D^{approx}_{MCT}$ shows a better agreement with $D_{sim}$. For system 
1 this discrepancy is so large that they are not plotted in the same graph. 
We find that the discrepancy reduces with the reduction of the strength of attractive force.  
We also find that the discrepancy is larger for smaller solutes. Thus the study reveals that for strong attractive interactions and smaller solutes the memory function for the self dynamic structure factor $\mathcal M^{s}_{qij}$  as given by Eq. \ref{fsmemory} is an approximation and should have contribution from other terms.

\subsection{ Levitation and radial distribution function}

Our analysis reveals that the non monotonic size dependence of diffusion is directly connected to the non monotonic behaviour of the 1st peak of the RDF. To get a deeper insight and also to understand the connection between levitation and structure of the liquid,  we compare the radial distribution functions for all the systems. This comparison is done for two solute sizes,  $\sigma_{22}=0.2439$, which falls in the normal region where smaller particles always diffuse faster and $\sigma_{22}=0.073$ which falls in the region where the smaller particles can have lower diffusion values compared to that for larger particles. The idea of this analysis is to understand how much the structural information can tell us about the dynamics in the system.  

\begin{figure}[h]
\centering
\includegraphics[width=12cm,height=6cm]{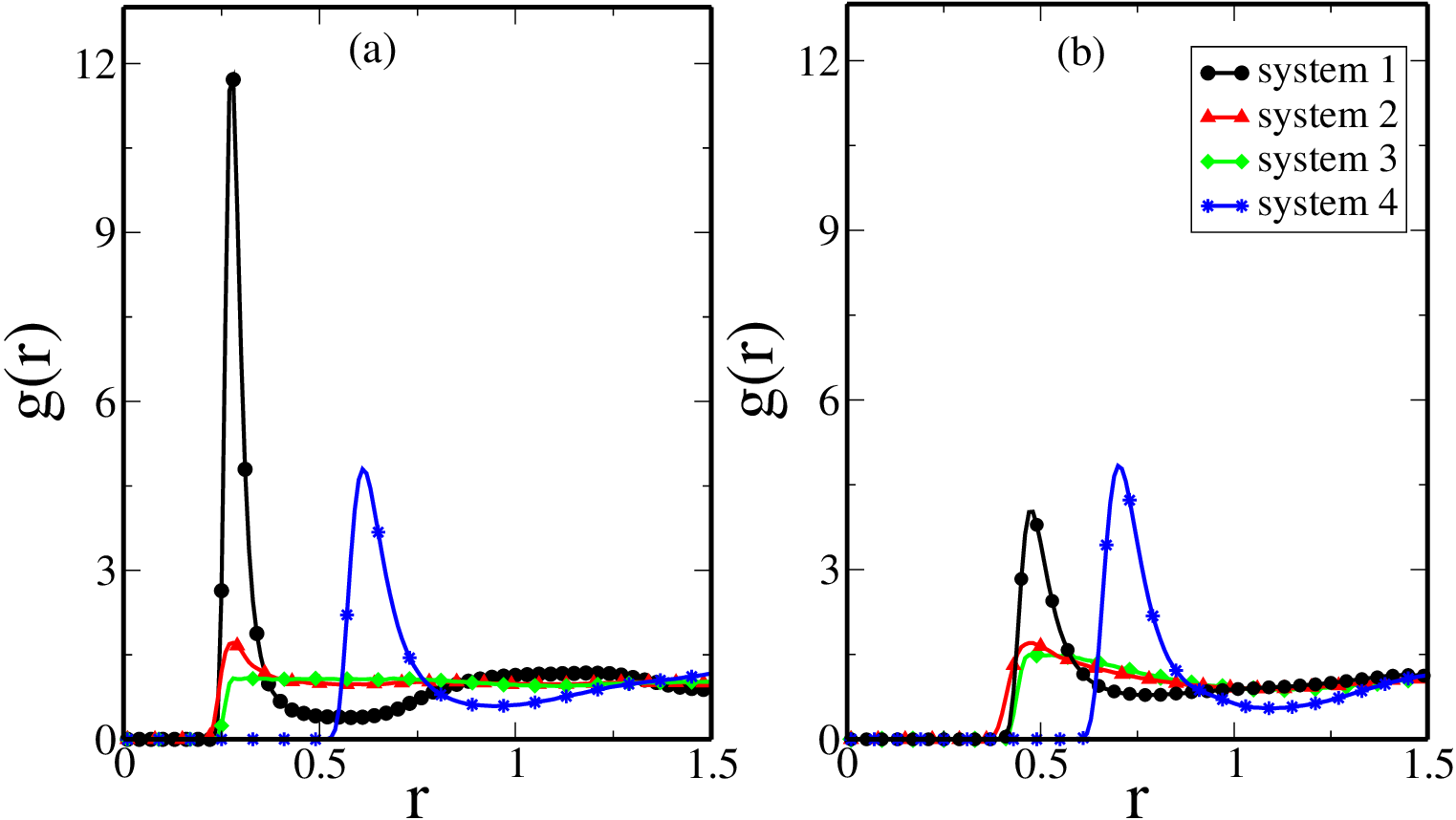}
\caption{Solute-solvent radial distribution function (RDF) for four different systems (1,2 3 and 4), where $\sigma_{22}$=0.073(a)  and $\sigma_{22}$=0.244 (b). For $\sigma_{22}$=0.073 the RDF in system 2 and 3 are diffused but in system 1 and 4 it has well-defined peaks with clear maxima and minima. For $\sigma_{22}$=0.244  in system 2 and 3 the peak values are diffused and in comparison to the small solute, the peak value for system 1 is much less. For system 4 since there is no levitation the peak values for both the sizes are similar.}
\label{fig7}
\end{figure}

 For non-anomalous systems the peak value of the RDF usually decreases with solute size (see Fig.\ref{fig5}b) and also as the attractive interaction between the solute-solvent is reduced. 
For system 2 and 3, for solute size $\sigma_{22}=0.2439$, the RDF has a diffused 1st peak as shown in Fig.\ref{fig7}b.  As the attraction between the solute and the solvent is increased, for system 1 we do see a well defined maxima but a weak signature of the minima is present and there is a flattening of the RDF after the minima.  When a solute levitates through the inter solvent pore the solute-solvent RDF is expected to have less structure which is clearly observed here.  For solute size 0.073, for system 2 and 3 the RDF is almost structure less due to the size effect, however for system 1 it has a well defined structure with proper maxima and minima.  Thus the RDF predicts that the solute remains 
closer to one solvent molecule and it cannot levitate through the transient solvent cage due to asymmetry in the interactions. 

A comparison between the RDFs for system 1 and 4 reveals that not only the solute size or the solute-solvent interaction potential but the solute-solvent approach distance also effects the peak value of the RDF and thus the dynamics. For the smaller solute the RDF in system 1 has a larger value of the 1st peak and for the intermediate solute the RDF in system 4 has a larger value of the 1st peak. 

Note that in system 2 although the diffusion does not show any non-monotonic behaviour but the RDF does (see Fig. \ref{fig5}a). Hence even in system 2 the intermediate size particle levitates.  
Historically it is shown that when the solvent forms a cage and the solute-solvent interaction is attractive then there is a force balance and thus solutes of certain sizes levitate \cite{deroune}\cite{santi_jpc98}. However, we believe that the levitation is rather dependent on the relative size of the solute with respect to the neck of solvent cage rather than the interaction between the solute-solvent. Whatever be the interaction potential, as long as it is isotropic, solutes of certain sizes will always take advantage of the curvature formed by the solvent molecules around them and in case of attractive interaction (be it weak or strong) there should be a small region where there is force cancellation and for the WCA potential there is a region depending on the solute size where the repulsive forces from all directions are absent. Since for the WCA potential the range of interaction is exactly where the first minima in the LJ potential is present so solutes of intermediate sizes which can levitate in presence of the strong LJ potential should also levitate in case of weak LJ or WCA potential.

The question that now arises is, if for system 2 and 3, the intermediate size solute levitates then why it does not show a diffusivity maximum. According to the Stokes-Einstein relationship, a smaller solute should diffuse faster. However as discussed earlier, solutes of certain sizes have a higher diffusion coefficient as they can levitate. When the effect of levitation exceeds the gain due to reduction of solute size then we see a non-monotonic size dependence of diffusion which happens in the case of system 1. However, as we lower the solute-solvent attractive interaction this gain due to levitation does not exceed the gain due to reduction of solute size and thus we do not see any non-monotonic size dependence of diffusion. This point will be studied in detail in future.

\section{Conclusion}

In this article we have studied using both analytical MCT and computer simulations the size dependence of diffusion of small solute molecules. This study is done in order to understand the non monotonic size dependence of diffusion of small solutes which was earlier found in simulation studies \cite{yashonath}. 
In the earlier simulation and experimental studies the diffusion of small solute particles in porous zeolite structures \cite{gh}
\cite{yashonath2010}, pure solvents\cite{yashonath} and ionic medium\cite{yasho196} have predicted that the non monotonic size dependence of diffusion is a generic phenomena, independent of the nature of the solvent.
Yashonath and co-workers have connected this behaviour to levitation effect [8-14]. They have found that when a solute size is approximately similar to the size of the zeolitic pore or in case of pure solvent, the size of the neck of the solvent cage then the solute particle feels a symmetric attraction from all directions and thus levitates through the pore and diffuses faster. However, a smaller solute gets closer to one wall of the pore or one solvent molecule and thus feels an asymmetric force and spends a longer time in the neck of the cage before diffusing out. This leads to a lower diffusion value for smaller solutes. 

In order to understand this phenomena we use both analytical MCT and simulation techniques. Study has been done for four different systems. In three systems (1, 2 and 3) the solute-solvent diameter is such chosen that interpenetration between the solute and the solvent is allowed. It is found that for system 1 which has strong attractive interaction the diffusion coefficient does show a non-monotonic size dependence of diffusion. For system 2, which has weak attractive interaction and system 3 which has only repulsive interaction there are no non-monotonic  behaviour as obtained from both analytical and simulation studies. 
This implies that systems with only strong attractive interactions show this anomalous behaviour. This finding is similar to the earlier reported simulation studies of sorbate diffusion in zeolites \cite{santi_jpc98}. 

To understand the effect of the solute-solvent interpenetration,  we have simulated a system which is strongly attractive but the solute-solvent diameter is calculated using LB rule. In this system again we do not find any non-monotonic size dependence. Note that, for a solute to levitate, the levitation parameter $\gamma$ needs to be close to but less than unity \cite{yashonath} \cite{yasho11}.  For any system where LB rule is followed the $\gamma$ value can never be less than unity as the solute-solvent diameters are always bigger than the neck of the solvent cage. Thus we conclude that for solute-solvent system to observe this anomalous size dependence, the attractive interaction between the solute and the solvent should be strong and also interpenetration between the solute-solvent pair should be allowed for the solute to explore the solvent cage. This solute-solvent interpenetration should not be
a hard criteria in a zeolitic medium as there are well defined pores. Thus in zeolitic medium as long as the solutes are smaller than the pore size, the non monotonic size dependence can be observed when there is strong interaction between the zeolitic wall and the solute particles. We also note that for system 1, 2 and 3 the primary contribution to the diffusion arises from the binary collisions. In a porous zeolite medium we expect no density relaxation of the medium. This should lead to a strong decoupling between the solute dynamics and the medium. Thus ideally the collisions between the solute and the zeolite structure should primarily determine the friction/diffusion. Hence, system 1,2 and 3 in some sense mimic the diffusion through a porous medium. 

The study further reveals that there is a strong one to one correlation between the non monotonic size dependence of diffusion and the non monotonic size dependence of the peak value of the solute-solvent RDF.

Although we use fully self consistent MCT but for smaller solutes we find that the self consistent calculation for the self dynamic structure factor predicts a long time tail which leads to larger friction value. The approximate self dynamic structure factor, using simulated mean square displacement and non-gaussian parameter does predict a fairly correct value for the diffusion. However for system 1 for the smallest two sizes there should be contribution from non hydrodynamic fast modes. Further, a more accurate memory function for the self dynamic structure factor is required when the solute sizes are smaller. 

In conclusion, for the first time the anomalous size dependence of diffusion 
for a non-polar solute-solvent system has been predicted using analytical MCT. The study reveals that this anomalous behaviour can be observed only when the solute-solvent interactions are strong and also when the solute can explore the solvent cage. The fact that the effect of a dynamic phenomena like levitation has its manifestation in the static RDF is quite encouraging. In future we would like to use analytical MCT to study diffusion in zeolitic medium.

\hspace{-0.7cm}{\large{\textbf{Acknowledgements}}}

This work was supported by the Department of Science and Technology (DST), India. MKN and AB thank DST for fellowship. SMB thanks Prof. S. Yashonath for discussions and for introducing her to this problem and Dr. A. Lele and Dr. G. Kumaraswamy for discussions.

\end{document}